\documentclass[12pt,a4paper]{article}
\usepackage[T1]{fontenc}
\usepackage{amssymb}
\usepackage{amsmath}
\usepackage{stmaryrd}
\usepackage{graphicx}
\usepackage{here}
\usepackage[margin=1in]{geometry}
\usepackage[colorlinks,allcolors=blue]{hyperref}
\usepackage{cite}
\usepackage{sectsty}
\allsectionsfont{\sffamily}

\newcommand{\be}{\begin{equation}}
\newcommand{\ee}{\end{equation}}
\newcommand{\ben}{\begin{enumerate}}
\newcommand{\een}{\end{enumerate}}
\newcommand{\bi}{\begin{itemize}}
\newcommand{\ei}{\end{itemize}}
\newcommand{\bea}{\begin{eqnarray}}
\newcommand{\eea}{\end{eqnarray}}
\newcommand{\nn}{\nonumber}

\newcommand{\eps}{\varepsilon}

\newcommand{\phii}{\varphi}

\newcommand{\der}{\partial}

\newcommand{\Diff}{\mathrm{Diff}^+(S^1)}
\newcommand{\Vect}{\mathrm{Vect}(S^1)}

\newcommand{\hatDiff}{\widehat{\mathrm{Diff}}^+(S^1)}
\newcommand{\hatVect}{\widehat{\mathrm{Vect}}(S^1)}
\newcommand{\BMS}{\mathrm{BMS}_3}
\newcommand{\bms}{\mathfrak{bms}_3}
\newcommand{\hatbms}{\widehat{\mathfrak{bms}}_3}

\newcommand{\lp}{\left(}
\newcommand{\rp}{\right)}

\newcommand{\calA}{{\cal A}}
\newcommand{\calD}{{\cal D}}
\newcommand{\calE}{{\cal E}}

\newcommand{\calH}{{\cal H}}

\newcommand{\calO}{{\cal O}}
\newcommand{\calR}{{\cal R}}

\newcommand{\calT}{{\cal T}}

\newcommand{\Ob}{{\cal O}_p}
\newcommand{\Ad}{\mathrm{Ad}}

\newcommand{\SL}{\mathrm{SL}(2,\mathbb{R})}
\newcommand{\PSL}{\mathrm{PSL}(2,\mathbb{R})}

\renewcommand{\sl}{\mathfrak{sl}}

\newcommand{\CC}{\mathbb{C}}

\newcommand{\RR}{\mathbb{R}}
\newcommand{\ZZ}{\mathbb{Z}}

\newcommand{\refeq}[1]{\stackrel{\text{(\ref{#1})}}{=}}

\begin{document}

\hrule
\begin{center}
\Large{\bfseries{\textsf{Characters of the BMS Group in Three Dimensions}}}
\end{center}
\hrule
~\\

\begin{center}
\large{\textsf{Blagoje Oblak$^{*}$}}
\end{center}
~\\

\begin{center}
\begin{minipage}{.9\textwidth}\small \it  \begin{center}
   Physique Th\'eorique et Math\'ematique \\ Universit\'e Libre de
   Bruxelles and International Solvay Institutes \\ Campus
   Plaine C.P. 231, B-1050 Bruxelles, Belgium
 \end{center}
\end{minipage}
\end{center}

\vspace{1cm}

\begin{center}
\begin{minipage}{.9\textwidth}
\begin{center}{\bfseries{\textsf{Abstract}}}\end{center}
Using the Frobenius formula, we evaluate characters associated with certain 
induced representations of the centrally extended BMS$_3$ group. This computation involves a 
functional integral over a coadjoint orbit of the Virasoro group; a delta function localizes the integral to 
a single point, allowing us to obtain an exact result. The latter is independent of the specific 
form of the functional measure, and holds for all values of the BMS$_3$ central charges and all values 
of the chosen mass and spin. It can also be recovered as a flat limit of Virasoro characters.
\end{minipage}
\end{center}

\vfill
\noindent
\mbox{}
\raisebox{-3\baselineskip}{%
  \parbox{\textwidth}{\mbox{}\hrulefill\\[-4pt]}}
{\scriptsize$^*$ Research Fellow of the Fund for Scientific Research-FNRS
Belgium. E-mail: boblak@ulb.ac.be}

\thispagestyle{empty}
\newpage

\textsf{\tableofcontents}
\thispagestyle{empty}

\newpage
\setcounter{page}{1}
\pagenumbering{arabic}
\section*{Introduction}
\addcontentsline{toc}{section}{Introduction}

The $\BMS$ group is the asymptotic symmetry group of three-dimensional Einstein gravity with a vanishing 
cosmological constant at null infinity 
\cite{Ashtekar,BarnichCompere,BarnichTroessaert}. It is an 
infinite-dimensional extension of the Poincar\'e group in three dimensions. On account of the semi-direct 
product structure of $\BMS$, it was recently argued in \cite{BarnichOblak02,BarnichOblak03} that 
its irreducible unitary representations are induced representations 
\cite{Wigner,Mackey,MackeyBook,MackeyInf,Barut,Corn}. The latter turn out to be classified by coadjoint 
orbits of the Virasoro group \cite{LazPan,Witten,Balog,Guieu}, which appear in this context as 
infinite-dimensional generalizations of the usual hyperboloidal orbits of the Poincar\'e group. Owing to the 
analogy between $\BMS$ and Poincar\'e, it is tempting to call ``supermomentum'' the 
infinite-dimensional vector that generalizes the usual Poincar\'e momentum. In particular, the orbits of 
generic constant supermomenta are diffeomorphic to $\Diff/S^1$ while the orbit of 
the $\BMS$ vacuum is the universal Teichm\"uller space $\Diff/\PSL$ \cite{Guieu,Nag}. The 
corresponding representations may be called ``$\BMS$ particles'', generalizing the notion of particle 
defined by Poincar\'e symmetry.\\

The purpose of this paper is to compute the characters of rotations and supertranslations in certain induced 
representations of the (centrally extended) $\BMS$ group. Specifically, we will focus on representations 
based on orbits of constant supermomenta, which include the $\BMS$ vacuum and massive $\BMS$ 
particles. We will compute these 
characters using the Frobenius formula \cite{Kirillov,FultonHarris,Joos}, which roughly states that the 
character of a 
group element in an induced representation is a certain integral over the corresponding orbit. Since the 
orbits relevant for 
$\BMS$ are infinite-dimensional Virasoro coadjoint orbits, the integral required by the 
Frobenius formula is in that case a functional one, {\it a priori} leading to two complications: the first 
is the definition of the functional integral measure, and the second is the practical 
calculation of the integral.\\

Remarkably, for non-zero rotation angle, both of these complications simply disappear. Indeed, the Frobenius 
formula contains a delta function that localizes the functional integral to a single point $-$ the unique 
point invariant under rotations on the orbit. Provided we find local coordinates on the orbit in a 
neighbourhood of that point, we can write down a local expression for the measure, up to an unknown 
prefactor. 
Since induced representations are independent 
(up to unitary equivalence) of the choice of a measure on the orbit, the contribution of the prefactor 
vanishes upon integrating the delta function. This provides an exact expression for the resulting 
character. In contrast to the 
Virasoro group, whose characters (in unitary representations) depend heavily on the values of the 
central charge and the highest weight 
\cite{DiFran,Blum,FeiginFuchs01,Wassermann01,Wassermann02}, the $\BMS$ characters derived here are valid for 
all values of the central charges and (almost) any 
value of the 
constant supermomentum on the selected orbit. Given that the $\BMS$ group can be seen as a high-energy, high 
central charge limit of two Virasoro groups, this simplification should not come as a surprise.\\

The paper is organized as follows. In section \ref{Poincare}, we briefly 
review the theory of induced representations for semi-direct products and the resulting 
Frobenius formula for characters, before applying these considerations to the Poincar\'e group in three 
dimensions. In section \ref{BMS}, we then use this technique to compute $\BMS$ 
characters and discuss their relation to standard Virasoro characters at large 
central 
charge. The 
conclusion is devoted to certain open issues and possible extensions of this work.

\section{Induced representations and the Frobenius formula}
\label{Poincare}

This section begins with a lightning review of the theory of induced representations (subsection 
\ref{indRep}), which is then used to derive the Frobenius formula for the corresponding characters 
(subsection \ref{Frobb}). Subsection \ref{charPoin} contains an application of this formula to massive 
representations of the Poincar\'e group in three dimensions. For the sake of generality, we will 
not assume that an invariant measure exists on each orbit, and we will therefore use quasi-invariant 
measures instead. Up to this slight difference, our notations and conventions match those of 
\cite{BarnichOblak02}. For the record, we will make no attempt at mathematical rigour. In particular, in this 
section, we will assume that all standard regularity assumptions for the objects involved (such as local 
compactness or separability) are satisfied. We refer to \cite{Barut} for details.

\subsection{Induced representations of semi-direct products}
\label{indRep}

Here we first define the notion of quasi-invariant measures (and introduce, in particular, the associated 
Radon-Nikodym derivative), before using it in the framework of induced representations.

\subsubsection*{Quasi-invariant measures}

Let $G$ be a topological group, acting continuously and transitively 
on a topological space $\calO$, with the action of 
$f\in G$ on $\calO$ denoted by 
$q\mapsto f\cdot q$ for all $q\in\calO$. Let $\mu$ be a (Borel) measure on $\calO$. We say that $\mu$ is 
{\it quasi-invariant} under the $G$ action if, for 
any $f\in G$, the measure $\mu_f$ defined by
\be
\int_{\calA}d\mu_f(q)\equiv\int_{f\cdot\calA}d\mu(q)\quad\text{for any measurable set }\calA\subseteq\calO
\nn
\ee
is equivalent to $\mu$, that is, if $\mu$ and $\mu_f$ have the same sets of measure zero. By 
virtue of the Radon-Nikodym theorem \cite{Barut,RudinRnC}, this amounts to saying that, 
for 
each $f\in G$, there exists a positive function $\rho_f$ on $\calO$, called the 
{\it Radon-Nikodym} derivative of $\mu_f$ with respect to $\mu$, such that $d\mu_f(q)=\rho_f(q)d\mu(q)$ for 
all $q\in\calO$. This 
relation is written as
\be
\rho_f(q)\equiv\frac{d\mu_f(q)}{d\mu(q)}=\frac{d\mu(f\cdot q)}{d\mu(q)}.
\label{RN}
\ee
Since $\mu_f$ is related to $\mu$ 
by a group action, the Radon-Nikodym derivative satisfies the following property:
\be
\rho_{fg}(q)=\rho_f(g\cdot q)\rho_g(q)\quad\forall\,f,g\in G,\;\forall\,q\in\calO.
\label{RNgroup}
\ee
If $\rho_f(q)=1$ for all $f\in G$ and any $q\in\calO$, 
we say that the measure $\mu$ is {\it invariant} under the action of $G$. For example, the 
usual Lebesgue measure $d^nx$ in $\RR^n$ is invariant under translations and rotations, but 
it is quasi-invariant under diffeomorphisms: under $f:x\mapsto f(x)$, the measure 
transforms into $d^nf(x)=|\det(\der f/\der x)|d^nx$, so the corresponding Radon-Nikodym derivative 
$\rho_f(x)=|\det(\der f/\der x)|$ is the absolute value of the
Jacobian determinant.\\

Although our presentation in the following pages will take into account the possibility of quasi-invariant 
measures, such subtleties will play no role once we turn to 
characters. Indeed, we will see that the Radon-Nikodym derivative does not contribute to the 
Frobenius formula ({\it cf.}~eq.~(\ref{Frobis}) below).

\subsubsection*{Induced representations}

The theory of representations of semi-direct products of the form $G\ltimes_{\sigma}A$, where $A$ is an
Abelian vector group, is well known\footnote{Henceforth we will systematically drop the 
subscript 
$\sigma$ in $G\ltimes_{\sigma}A$.}. Under suitable regularity assumptions, it turns out that 
all irreducible 
unitary representations of such groups are so-called induced representations 
\cite{Mackey,MackeyBook,MackeyInf,Barut}.\\

We begin by fixing some notation \cite{BarnichOblak02}. Let $A^*$ be the dual of $A$. For 
any $p\in A^*$, define the action of $f\in G$ on $p$ by $\langle f\cdot p,\alpha\rangle\equiv\langle 
p,\sigma_{f^{-1}}\alpha\rangle$ for all $\alpha\in A$. Let then
\be
\calO_p
\equiv
\left\{f\cdot p|f\in G\right\}\subset A^*
\label{orb}
\ee
be the orbit of $p\in A^*$ under this action, and let
\be
G_p
\equiv
\left\{f\in G|f\cdot p=p\right\}
\nn
\ee
be the corresponding little group. We will assume that there exists a measure $\mu$ on $\calO_p$ 
that is quasi-invariant under the action of $G$. Finally, seeing the group $G$ as a principal $G_p$-bundle 
over $\calO_p$, introduce a continuous section
\be
g:{\cal O}_p\rightarrow G:q\mapsto g_q
\label{stdBoost}
\ee
such that $g_q\cdot p=q$ for all $q\in{\cal O}_p$. In writing this, we are assuming that $g$ is a global 
section, {\it i.e.}~that the $G_p$-bundle $G\rightarrow\calO_p$ is trivial. This property is not true in 
general but it will hold in all cases of interest below, so we will stick to it.\\

Consider now a unitary representation
$\calR:G_p\rightarrow\mathrm{GL}\lp\calE\rp:f\mapsto\calR[f]$ of a
given little group $G_p$ in a complex Hilbert space $\calE$,
equipped with a scalar product $(\cdot|\cdot)$. Let also $\calH$ denote the Hilbert space of ``wavefunctions''
$\Psi:\Ob\rightarrow\calE$ that are square-integrable with respect to the measure $\mu$, endowed with the 
scalar
product
\be
\langle\Phi|\Psi\rangle
\equiv
\int_{\Ob}d\mu(q)\left(\Phi(q)|\Psi(q)\right).
\label{scal}
\ee
The {\it induced representation} $\calT$ associated with $\calR$ then acts in $\calH$ according to
\cite{Barut,Corn}
\be
\lp\calT\left[\lp f,\alpha\rp\right]\Psi\rp(q)
\equiv
\left[\rho_{f^{-1}}(q)\right]^{1/2}
e^{i\langle q,\alpha\rangle}\calR\left[g_q^{-1}fg_{f^{-1}\cdot q}\right]\Psi\lp
f^{-1}\cdot q\rp
\label{calt}
\ee
for all $q\in\calO_p$ and any 
wavefunction $\Psi$ in $\calH$; here $\rho$ denotes the Radon-Nikodym derivative of the measure $\mu$, as 
defined in (\ref{RN}). It is easily verified, using 
the above definitions (and in particular relation (\ref{RNgroup})), that this expression indeed defines a 
representation of $G\ltimes A$. In fact, this would be true even without the prefactor 
$\left[\rho_{f^{-1}}(q)\right]^{1/2}$; the latter is needed, however, 
to make 
this representation unitary\footnote{In 
\cite{BarnichOblak02} we limited ourselves to invariant measures, so the Radon-Nikodym derivative 
never showed up.} with respect to the scalar product (\ref{scal}). Note that different 
choices of quasi-invariant measures lead to 
unitarily equivalent induced representations: if $\mu_1$ and $\mu_2$ are two such measures on $\calO_p$, 
each defining a scalar product of the form (\ref{scal}) in the Hilbert spaces of $\calE$-valued 
square integrable wavefunctions $\calH_1$ and $\calH_2$ (respectively), then the map
\be
U:\calH_1\rightarrow\calH_2:\Psi\mapsto U\Psi
\quad \text{with}\quad
\left(U\Psi\right)(q)\equiv\left[\frac{d\mu_1(q)}{d\mu_2(q)}\right]^{1/2}\Psi(q)
\label{UU}
\ee
is an isometry that intertwines the induced representations $\calT_1$ and $\calT_2$. (Here $d\mu_1/d\mu_2$ 
is the Radon-Nikodym derivative of $\mu_1$ with respect to $\mu_2$.)\\

When the vector group $A$ is seen as a group of translations, its dual $A^*$ consists 
of ``momenta''. Then the $G$-orbits defined by (\ref{orb}) are the usual momentum orbits, familiar 
for 
instance from the Poincar\'e group. In that case the map (\ref{stdBoost}) defines the ``standard 
boost'' $g_q$ for each $q$ on the orbit, and the measure $\mu$ is a Lorentz-invariant momentum measure on 
$\calO_p$. 
For example, for the orbit of a particle with mass $m$ living in $(n+1)$-dimensional Minkowski 
space-time, one has $d\mu(q)=d^n{\bf q}/\sqrt{m^2+{\bf q}^2}$, where ${\bf q}$ denotes the spatial 
momentum. In that context, the space $\calE$ of the representation $\calR$ of the little group 
is the space of ``internal'' degrees of freedom, and the label specifying $\calR$ 
is called ``spin''. From now on, we will 
freely use the terminology of ``momenta'' and ``spin'' for any semi-direct 
product $G\ltimes A$ (with $A$ a vector group), as this terminology is also appropriate for the $\BMS$ 
group.\\

Formula (\ref{calt}) can be conveniently re-expressed in terms of a basis of delta 
functions, physically
representing ``plane waves'' or particles with definite momentum. To define this basis, introduce the 
Dirac delta distribution $\delta_{\mu}$ associated with 
the measure $\mu$, such that
\be
\int_{\calO_p}d\mu(q)\delta_{\mu}(k,q)\varphi(q)=\varphi(k)
\label{testFct}
\ee
for any test function $\varphi$ on $\calO_p$ and any $k\in\calO_p$. Since in general $\mu$ transforms 
non-trivially under the $G$ 
action, the corresponding delta function transforms as
\be
\delta_{\mu}(f\cdot k,f\cdot q)=\left[\rho_f(q)\right]^{-1}\delta_{\mu}(k,q).
\nn
\ee
Now, if $\{e_m|m=1,2,3,...\}$ is an orthonormal basis of
$\cal E$, define the wavefunction of definite momentum $k$ and polarization $m$ to be
\be
\Psi_{k,m}(q)
\equiv
\delta_{\mu}(k,q)e_m.
\label{1p}
\ee
The scalar product (\ref{scal}) of such wavefunctions is
\be
\left<\Psi_{k,m}|\Psi_{q,n}\right> =
\delta_{mn}\delta_{\mu}(k,q).
\label{scalPlane}
\ee
In other words, the set $\left\{\Psi_{k,m}|k\in\calO_p,\,m=1,2,3,...\right\}$ forms an orthonormal basis of 
$\calH$. In this basis, the action (\ref{calt}) of the induced representation becomes
\be
\calT\left[\lp f,\alpha\rp\right]\Psi_{k,m}
=
\left[\rho_f(k)\right]^{1/2}
e^{i\langle {f}\cdot k,\alpha\rangle}
\left(\calR\left[g_{f\cdot k}^{-1}\,fg_k\right]\right)_{nm}\Psi_{f\cdot k,n},
\label{induced1p}
\ee
where summation over repeated indices is implicit, and where $\left(\calR\left[g_{f\cdot 
k}^{-1}\,fg_k\right]\right)_{nm}$ denotes the matrix element of $\calR\left[g_{f\cdot 
k}^{-1}\,fg_k\right]$ between the vectors $e_n$ and $e_m$.

\subsection{Characters: the Frobenius formula}
\label{Frobb}

The character of a representation is the map that associates, with each group element, the trace of the 
operator that represents it. In the present case, we want to compute the character of the 
induced representation $\calT$,
\be
\chi:G\ltimes A\rightarrow\CC:(f,\alpha)\mapsto\text{Tr}\big(\calT[(f,\alpha)]\big).
\nn
\ee
Since the carrier space of $\calT$ is a Hilbert space $\calH$, the trace of $\calT\left[(f,\alpha)\right]$ 
can be written as a sum of scalar products between vectors of an orthonormal basis of $\calH$ and their 
images under $\calT\left[(f,\alpha)\right]$. For the representation based on $\calO_p$, with spin $\calR$, a 
convenient orthonormal basis is provided by plane waves $\Psi_{k,n}$, 
as defined in (\ref{1p}). The ``sum'' over scalar products then 
becomes an integral over $\calO_p$ \cite{Joos}:
\be
\text{Tr}\big(\calT[(f,\alpha)]\big)
\equiv
\chi[(f,\alpha)]
=
\int_{\calO_p}d\mu(k)\sum_{n=1}^{+\infty}\big<\Psi_{k,n}\big|\calT[(f,\alpha)]\Psi_{k,n}\big>,
\nn
\ee
where $\mu$ is the quasi-invariant measure on $\calO_p$ used to define $\calT$. By virtue of the action 
(\ref{induced1p}) of the 
induced representation on plane waves, this
character can be expressed as
\be
\chi[(f,\alpha)]
=
\int_{\calO_p}d\mu(k)
\left[\rho_f(k)\right]^{1/2}
e^{i\langle f\cdot k,\alpha\rangle}
\sum_{m,n=1}^{+\infty}
\left(\calR\left[g_{f\cdot k}^{-1}\,fg_k\right]\right)_{mn}
\big<\Psi_{k,n}\big|\Psi_{f\cdot k,m}\big>.
\nn
\ee
The normalization
(\ref{scalPlane}) then allows us to rewrite this as
\bea
\chi[(f,\alpha)]
& = &
\int_{\calO_p}d\mu(k)
\left[\rho_f(k)\right]^{1/2}
\delta_{\mu}\left(k,f\cdot k\right)
e^{i\langle f\cdot k,\alpha\rangle}
\sum_{n=1}^{+\infty}
\left(\calR\left[g_{f\cdot k}^{-1}\,fg_k\right]\right)_{nn}\nn\\
& = &
\int_{\calO_p}d\mu(k)
\left[\rho_f(k)\right]^{1/2}
\delta_{\mu}\left(k,f\cdot k\right)
e^{i\langle k,\alpha\rangle}
\chi_{\calR}\left[g_k^{-1}fg_k\right],\nn
\eea
where $\chi_{\calR}$ denotes the character of $\calR$. The delta function restricts the integration to the 
subset of $\calO_p$ consisting of points $k$ such that $f\cdot k=k$. In particular, this allows us to set the 
Radon-Nikodym derivative to one, since $\rho_f(k)=d\mu(f\cdot k)/d\mu(k)=d\mu(k)/d\mu(k)=1$ on that subset. 
Our final formula for the character of $\calT$ is thus
\be
\chi[(f,\alpha)]
=
\int_{\calO_p}d\mu(k)
\delta_{\mu}\left(k,f\cdot k\right)
e^{i\langle k,\alpha\rangle}
\chi_{\calR}\left[g_k^{-1}fg_k\right].
\label{Frobis}
\ee
This is the {\it Frobenius formula} for characters of induced representations of $G\ltimes A$ \cite{Joos}.\\

From now on, we will no longer need to take care of Radon-Nikodym derivatives, since they do not 
contribute to 
characters. In fact, we should have expected this simplification: even though different quasi-invariant 
measures have different Radon-Nikodym derivatives in general, they lead to unitarily equivalent induced 
representations ({\it cf.}~the discussion around (\ref{UU})). Since the characters of equivalent 
representations 
are identical, the quantity (\ref{Frobis}) cannot depend on the measure, and must therefore be independent of 
its Radon-Nikodym derivative. Note also that (\ref{Frobis}) is a {\it class function}, as it 
should: it depends only on the conjugacy class of the group element $(f,\alpha)$ at which it is evaluated.\\

The Frobenius formula (\ref{Frobis}) states, roughly speaking, that the character  $\calT$ is a ``sum'' of 
characters of $\calR$ \cite{Kirillov,FultonHarris}. More precisely, we may recognize
\be
e^{i\langle k,\alpha\rangle}
\chi_{\calR}\left[g_k^{-1}fg_k\right]
\label{makeSense}
\ee
as the character of an irreducible unitary representation of $G_p\ltimes A$, evaluated at 
$(g_k^{-1}fg_k,\alpha)$. If $g_k^{-1}fg_k$ does not belong to $G_p$, expression (\ref{makeSense}) as such 
does not make 
sense, but the delta function $\delta_{\mu}\left(k,f\cdot k\right)$ ensures that such $k$'s do not contribute 
to the integral: whenever $k\neq f\cdot k$, the integrand of (\ref{Frobis}) vanishes. (By contrast, when 
$k=f\cdot k$, then $g_k^{-1}fg_k=g_{f\cdot k}^{-1}\,fg_k$ automatically belongs to $G_p$.) More generally, 
when 
$f$ is not conjugate to an element of $G_p$, $\chi[(f,\alpha)]$ vanishes.

\subsection{Characters of the Poincar\'e group in three dimensions}
\label{charPoin}

The characters of induced representations of the Poincar\'e group in four dimensions were computed in 
\cite{Joos,Fuchs}. Here we apply formula (\ref{Frobis}) to perform an analogous computation for the 
Poincar\'e group in three dimensions. For simplicity, we will focus on the case of a massive particle. Our 
goal is both to give a concrete illustration of the Frobenius 
formula and to use this example later, as a guide for the characters of $\BMS$.\\

A relativistic massive particle (in three space-time dimensions) is an induced representation of the 
Poincar\'e group 
$\SL\ltimes_{\text{Ad}}\sl(2,\RR)$ based on the orbit of momenta $q=(q_0,q_1,q_2)$ with positive energy 
satisfying
\be
-(q_0)^2+(q_1)^2+(q_2)^2\equiv q_{\mu}q^{\mu}=-m^2,
\label{q2}
\ee
where $m$ is a positive constant. We take the orbit representative to be $p=(m,0,0)$, so the orbit $\calO_p$
is the one-sheeted hyperboloid in momentum space going through $p$ and defined by (\ref{q2}). The
corresponding little group is $\mathrm{U}(1)$, and therefore the spin of the particle is {\it a priori} 
an
integer. However, the fundamental group of the Poincar\'e group in three dimensions is $\ZZ$ so that 
projective representations need to be taken into account \cite{Weinberg}. Hence the spin may actually take 
any real value and an
appropriate (irreducible, unitary, projective) representation $\calR$ of the little group is given by
\be
\calR[\text{rotation by $\theta$}]
=
e^{ij\theta},
\nn
\ee
where $j$ is any real number. The corresponding space $\calE$ is just $\CC$.\\

As explained above, the character (\ref{Frobis}) vanishes whenever $f$ is not conjugate to an element of the 
little group. In the present case, this means that $\chi[(f,\alpha)]=0$ whenever $f\in\SL$ is not conjugate 
to an element of the $\mathrm{U}(1)$ subgroup of $\SL$ leaving $p=(m,0,0)$ fixed. Thus, in order to obtain a 
non-trivial result, we must assume that $f$ is conjugate to a rotation by some (possibly vanishing) 
angle $\text{Rot}(f)\equiv\theta$. In that case, for any $k$ in $\calO_p$ such that $f\cdot k=k$, the 
character $\chi_{\calR}$ appearing in (\ref{Frobis}) takes the value 
$\chi_{\calR}\left[g_k^{-1}fg_k\right]=e^{ij\theta}$. Other $k$'s do not contribute to the integral 
(\ref{Frobis}) because of the delta function $\delta_{\mu}\left(k,f\cdot k\right)$, so we are free to pull 
$\chi_{\calR}$ out of the integral:
\be
\chi[(f,\alpha)]
=
e^{ij\theta}
\int_{\calO_p}d\mu(k)
\delta_{\mu}\left(k,f\cdot k\right)
e^{i\langle k,\alpha\rangle}.
\label{pchar}
\ee
To compute the integral as such, we may choose the momentum measure $\mu$ to be the Lorentz-invariant volume 
form
\be
d\mu(q)=\frac{d^2{\bf q}}{\sqrt{m^2+{\bf q}^2}}=\frac{dq_1dq_2}{\sqrt{m^2+q_1^2+q_2^2}},
\label{measLor}
\ee
the associated delta function being
\be
\delta_{\mu}(p,q)
=
\sqrt{m^2+q_1^2+q_2^2}\;\delta(p_1-q_1)\delta(p_2-q_2),
\label{deltaLor}
\ee
in accordance with the definition (\ref{testFct}). The delta functions appearing on the right-hand 
side are the standard Dirac delta functions on the real line. Plugging these expressions in (\ref{pchar}), 
the 
prefactors involving $\sqrt{m^2+{\bf q}^2}$ cancel and we find
\be
\chi[(f,\alpha)]
=
e^{ij\theta}
\int_{\RR^2}dk_1dk_2e^{i\langle k,\alpha\rangle}
\delta\left(k_1-[f\cdot k]_1\right)
\delta\left(k_2-[f\cdot k]_2\right),
\label{integral}
\ee
where the subscripts $1$ and $2$ denote the corresponding spatial components. Since by assumption $f$ is 
conjugate to a rotation (by an angle $\theta$), and since the character is a class function, we are free 
to replace $f$ by a pure rotation $f_{\theta}$ in this equation:
\be
\chi[(f,\alpha)]
=
e^{ij\theta}
\int_{\RR^2}dk_1dk_2e^{i\langle k,\alpha\rangle}
\delta\left(k_1-k_1\cos\theta+k_2\sin\theta\right)
\delta\left(k_2-k_1\sin\theta-k_2\cos\theta\right).
\nn
\ee
At this point we must consider separately two distinct cases:
\begin{itemize}
\item If $\theta\neq0$, the only
point $k$ on $\calO_p$ such that $f_{\theta}\cdot k=k$ is $p=(m,0,0)$. The delta function then ``localizes''
the integral, giving rise to the character
\be
\chi[(f,\alpha)]
=
e^{ij\theta}e^{im\alpha^0}
\Big|\text{det}\big(\mathbb{I}-f_{\theta}\big)\Big|^{-1},
\label{intm1}
\ee
where $\alpha^0$ denotes the time component of the translation vector $\alpha$ $-$ the only component 
that survives after integrating the delta function. The determinant is
\be
\text{det}\big(\mathbb{I}-f_{\theta}\big)
=
\left|
\begin{matrix}
1-\cos\theta & \sin\theta \\ -\sin\theta & 1-\cos\theta
\end{matrix}
\right|
=
4\sin^2(\theta/2),
\nn
\ee
so that
\be
\chi[(f,\alpha)]
=
e^{ij\theta}e^{im\alpha^0}\frac{1}{4\sin^2(\theta/2)}
\quad
\text{if $\text{Rot}(f)=\theta\neq0$}.
\label{charrot}
\ee
\item If $f=e$ is the identity, $\theta=0$. In that case, the argument of
the delta function in (\ref{integral}) is always zero; we interpret $\delta^{(2)}(0)$ as the spatial volume 
$V$ of
the system, up to factors of $2\pi$:
\be
\delta(k_1-k_1)\delta(k_2-k_2)=\frac{V}{(2\pi)^2}.
\label{deltaZero}
\ee
We thus find
\be
\chi[(e,\alpha)]
=
\frac{V}{(2\pi)^2}
\int_{\RR^2}dk_1dk_2e^{i\langle k,\alpha\rangle}.
\nn
\ee
We will not evaluate this integral for arbitrary $\alpha$. In the special case 
where $\alpha$ is a pure time translation $\alpha_t$ (so that $\alpha^0\equiv t$ is the only non-vanishing 
component of $\alpha$), we obtain
\be
\chi[(e,\alpha_t)]
=
\frac{V}{(2\pi)^2}
\int_{\RR^2}dk_1dk_2
e^{it\sqrt{m^2+k_1^2+k_2^2}}
=
\frac{V}{2\pi}
\int_{m}^{+\infty}udue^{itu}.
\label{timeInt}
\ee
Let us take $t>0$ for definiteness. Then, adding a small imaginary part $i\epsilon$ to $t$ to ensure 
convergence, the character of a pure time translation becomes
\be
\chi[(e,\alpha_t)]
=
\frac{V}{2\pi}
\int_{m}^{+\infty}udue^{i(t+i\epsilon)u}
=
-\frac{V}{2\pi t^2}
\left(1-imt\right)e^{imt}.
\label{timeChar}
\ee
\end{itemize}

So much for the Poincar\'e group in three dimensions. Before going further, let us stress a point that will 
be crucial once we turn to $\BMS$: in the steps 
leading from (\ref{pchar}) to (\ref{intm1}), the actual form of the measure $\mu$ mattered very little. 
Indeed, the function multiplying $dq_1dq_2$ in (\ref{measLor}) cancelled the prefactor of the delta function 
(\ref{deltaLor}), so the only important information was that the volume form on the orbit must 
be proportional to $dq_1dq_2$. Any other quasi-invariant momentum measure satisfying this basic requirement 
would have 
given the same 
result (\ref{intm1}) for the character, since the prefactor of $dq_1dq_2$ in that measure would have been 
cancelled by the inverse prefactor appearing in the corresponding delta function. This cancellation 
is ensured by the very definition (\ref{testFct}) of the delta function $\delta_{\mu}$. From a 
group-theoretic viewpoint, this simplification was to be expected: as 
already mentioned, two induced representations built using two different quasi-invariant measures on the 
orbit 
are (unitarily) equivalent. Since characters of equivalent representations are identical, 
they cannot depend on the choice of a measure.

\section{Characters of the {\bf $\text{BMS}_3$} group}
\label{BMS}

We now apply the procedure described above to the (centrally extended) $\BMS$ group: we begin by reviewing 
the structure of $\BMS$ 
(subsection \ref{indBMS}), before applying the Frobenius formula to the computation of characters associated 
with massive $\BMS$ particles (subsection \ref{charrBMS}). In subsection \ref{subsecCompChar}, we then 
establish the relation between these characters and those of highest weight representations of the Virasoro 
algebra. Finally, subsection \ref{ssecVac} is devoted to the character of the $\BMS$ vacuum representation. 
For more details on induced
representations of $\BMS$ and the associated terminology, we refer to \cite{BarnichOblak02,BarnichOblak03}. 
As before, our presentation will not be mathematically rigorous. In particular, we will assume that the 
theory of induced representations, as outlined in subsection \ref{indRep}, is applicable to the $\BMS$ group 
even though the latter is infinite-dimensional.

\subsection{Induced representations of the {\bf $\text{BMS}_3$} group}
\label{indBMS}

\subsubsection*{The {\bf $\text{BMS}_3$} group}

Let us first collect some background material on the $\BMS$ group and its representations. Recall the
definition \cite{BarnichOblak02}
\begin{center}
\begin{tabular}{c}
$\BMS\equiv\;\Diff\ltimes_{\Ad}\Vect_{\text{ab}}$\\
$\quad\quad\quad\swarrow\quad\quad\quad\quad\quad\quad\quad\;\searrow$\\
$\quad\quad\quad\quad$``superrotations''$\;\quad\quad\quad\quad\quad\;$``supertranslations''
\end{tabular}
\end{center}
where $\Diff$ denotes the group of (orientation-preserving) diffeomorphisms of the circle while
$\Vect_{\text{ab}}$ denotes the Abelian additive group of vector fields on the circle. The same structure
remains valid in the centrally extended case, with $\Diff$ and $\Vect$ replaced by the Virasoro group
$\hatDiff$ and its algebra, $\hatVect$; their semi-direct product is the centrally extended $\BMS$ group.
In the following pages we will use the same notation for the $\BMS$ group and its central extension, as the 
context should make it clear enough which group we are dealing with.\\

The dual of the group of supertranslations ({\it i.e.}~the space $A^*$ in the notations of subsection 
\ref{indRep}) is the dual space of the Virasoro algebra and consists of {\it supermomentum} vectors 
representing an infinite-dimensional generalization of the usual Poincar\'e momentum. In practice, in terms 
of a $2\pi$-periodic angular coordinate $\phii$ on the circle, a 
supermomentum $p$ is a quadratic density $p(\phii)d\phii^2$ paired with supertranslations according 
to\footnote{The definition (\ref{pairing}) differs by a factor of $2\pi$ from the convention used in 
\cite{BarnichOblak02,BarnichOblak03}. This ensures that the zeroth Fourier mode of $p(\phii)$ coincides with 
the associated energy.}
\be
\langle p,\alpha\rangle
\equiv
\frac{1}{2\pi}\int_0^{2\pi}d\phii\,p(\phii)\alpha(\phii)
\quad\forall\;\alpha(\phii)\frac{d}{d\phii}\in\Vect_{\text{ab}}.
\label{pairing}
\ee
Under the action of superrotations, supermomenta
transform according to the coadjoint representation of the Virasoro group. Explicitly, if $f:\phii\mapsto 
f(\phii)$ is a diffeomorphism of the circle\footnote{We use a slight abuse of notation in describing a 
diffeomorphism of the circle by one of its lifts in the universal cover of $\Diff$, {\it i.e.}~by a 
diffeomorphism of $\RR$ that satisfies $f(\phii+2\pi)=f(\phii)+2\pi$.}, this action is given by
\be
\left.\left(f\cdot p\right)\right|_{f(\phii)}
=
\frac{1}{(f'(\phii))^2}\left[p(\phii)+\frac{c_2}{12}S[f](\phii)\right].
\label{coadVir}
\ee
Here the prime denotes differentiation with respect to $\phii$, while
\be
S[f](\phii)=\frac{f'''(\phii)}{f'(\phii)}-\frac{3}{2}\left(\frac{f''(\phii)}{f'(\phii)}\right)^2
\nn
\ee
is the Schwarzian derivative of $f$ at $\phii$, and 
$c_2$ is the (dual of the) central charge pairing generators of superrotations with generators of 
supertranslations in the $\hatbms$ 
algebra (see 
expression (\ref{bms}) below). Upon taking an infinitesimal diffeomorphism $f(\phii)=\phii-\eps X(\phii)$, 
where the vector field $X(\phii)\frac{d}{d\phii}$ should be understood as an element of the Lie algebra of 
$\Diff$, 
the transformation law (\ref{coadVir}) yields (to first order in $\eps$)
\be
(f\cdot p)(\phii)-p(\phii)\equiv\eps\delta_Xp(\phii)
\quad\text{with}\quad
\delta_Xp=Xp'+2X'p-\frac{c_2}{12}X'''.
\label{cake}
\ee

In the context of three-dimensional asymptotically flat Einstein gravity, $p(\phii)$ is the Bondi mass 
aspect \cite{BarnichOblak03} transforming under asymptotic superrotations according to (\ref{coadVir}), and 
$c_2$ takes the value $3/G$, where $G$ denotes Newton's 
constant \cite{BarnichCompere}. As the 
subscript ``$2$'' indicates, there is also a central charge $c_1$, pairing superrotations with themselves as 
in the usual Virasoro algebra. However, $c_1$ plays virtually no role for induced representations (except for 
those based on the orbit of $p=0$ and $c_2=0$) and vanishes in Einstein gravity, so $c_1$ 
will almost never appear in what follows.\\

The $\BMS$ group being a semi-direct product, its (irreducible) unitary
representations are expected to be induced from those of its little groups, in the sense of subsection 
\ref{indRep}. The relevant orbits $\calO_p$ then are coadjoint orbits of the Virasoro group 
\cite{LazPan,Witten,Balog,Guieu}. The little group of a generic orbit is Abelian so the corresponding 
internal space $\calE$ is simply $\CC$, and the Hilbert space $\calH$ of the associated induced 
representations should be a space of complex-valued, square-integrable wavefunctionals on the orbit. (We say 
``wavefunctionals'' instead of ``wavefunctions'' to stress the fact that the orbits are, in this case, spaces 
of functions.) But in order to define what ``square-integrable'' means, we need a supermomentum measure on 
the orbit.

\subsubsection*{The question of the measure}

The problem of defining quasi-invariant measures on Virasoro coadjoint orbits is well known; see {\it 
e.g.}~\cite{AiraultMal,AiraultFrench,Dai}. In that context, a result that is especially relevant for our 
purposes is the theorem due to Shavgulidze 
\cite{ShavgulidzeTheOriginal,ShavgulidzeBis,ShavgulidzeFR,Shavgulidze} (see also 
\cite{Bogachev,Shimomura}) which states that the group of $C^k$ diffeomorphisms of any compact manifold can 
be 
endowed with a Borel measure that is quasi-invariant under the action of $C^{k+\ell}$ diffeomorphisms (here 
$k$ and $\ell$ are any two positive integers). This suggests that there exist quasi-invariant measures on 
$\Diff$, which in turn should provide quasi-invariant measures on a Virasoro coadjoint orbit $\calO_p$ 
with compact 
little group; indeed, if $\bar\mu$ is a quasi-invariant 
measure on $\Diff$ and if $\pi:\Diff\rightarrow\calO_p$ is the natural projection, we can define a 
quasi-invariant measure $\mu$ on 
$\calO_p$ by $\mu(\calA)\equiv\bar\mu\left(\pi^{-1}(\calA)\right)$ for any measurable subset 
$\calA\subseteq\calO_p$. When the little group is non-compact, however, this naive procedure may break down.\\

We will not dwell on such questions here and we will not attempt to make our considerations mathematically 
precise. Instead, our point of view will be a practical one: the supermomentum measures that we need are 
functional integral 
measures on certain (Fr\'echet) manifolds consisting of functions on the circle. Such measures are 
encountered on a daily basis in quantum mechanics, statistical 
physics and field theory. Provided one is willing to define Hilbert spaces of square-integrable functions 
with the help of functional 
measures, their use in induced representations of $\BMS$ 
is no more controversial than in quantum physics. Our hope is that techniques similar to those that led 
to a rigorous construction of path integral measures can be adapted to 
quasi-invariant 
measures on Virasoro coadjoint orbits. In the following pages, we will rely on 
this assumption to justify the use of functional integrals when computing characters.

\subsection{Characters of massive {\bf $\text{BMS}_3$} particles}
\label{charrBMS}

In this subsection we consider the induced representation based on the orbit 
$\calO_p$ of a constant 
supermomentum $p=m-c_2/24$ (with $c_2>0$), representing a $\BMS$ particle with 
rest 
mass $m>0$ normalized with respect to the vacuum. The condition $m>0$ ensures that 
energy is bounded from 
below on the orbit \cite{Balog}. We will denote by $j\in\RR$ the spin of 
the (projective) representation of the 
corresponding little group $\mathrm{U}(1)$ of rigid rotations of the circle. Our goal is to compute the 
character of this 
representation using the Frobenius formula (\ref{Frobis}).\\

Let us pick $(f,\alpha)\in\BMS$ and call $\chi_{m,j}[(f,\alpha)]$ the corresponding character. As 
explained earlier, the latter vanishes if $f$ is not conjugate to an element of the little 
group. To obtain a non-trivial result in the present case, we must therefore assume that $f$ is conjugate to 
a pure rotation, the angle of which is 
given by the Poincar\'e rotation number of $f$ \cite{Guieu}:
\be
\text{Rot}(f)=\lim_{n\rightarrow+\infty}\frac{f^n(\phii)-\phii}{n}\equiv\theta.
\label{BigStar}
\ee
Here $\phii\in[0,2\pi)$ is arbitrary and $f$ is seen as a diffeomorphism of $\RR$ satisfying 
$f(\phii+2\pi)=f(\phii)+2\pi$, with $f^n\equiv\underbrace{f\circ 
f\circ...\circ f}_{n\text{ times}}$. As in subsection \ref{charPoin}, the character $\chi_{\calR}$ of the 
little group can then be pulled out of the integral (\ref{Frobis}), producing an overall constant factor 
$e^{ij\theta}$. The character $\chi_{m,j}[(f,\alpha)]$ thus reduces to (\ref{pchar}), except that now the 
integral is taken over the (infinite-dimensional) Virasoro coadjoint
orbit $\calO_p$ and that $\langle k,\alpha\rangle$ denotes the pairing (\ref{pairing}) rather than a 
finite-dimensional scalar product. Since the character 
is a class function, we are free to replace $f$ by a pure rotation $f_{\theta}$ inside the integral. 
Depending on whether $\theta$ vanishes or not, we are then faced with two completely different problems.\\

If $\theta=0$ so that $f$ is actually the identity, the term $\delta_{\mu}(k,f\cdot 
k)=\delta_{\mu}(k,k)$ leads to an infrared divergence analogous to (\ref{deltaZero}). Assuming that we 
have regularized this divergence somehow, we are left with the task of computing a genuine path integral 
$-$ we must integrate the functional $\exp[i\langle k,\alpha\rangle]$ over all quadratic densities $k$ 
belonging to $\calO_p$. We shall not attempt to perform this computation here, only briefly returning to this 
issue in the conclusion of this work.\\

A radically different situation occurs if $\theta\neq0$. In that case, the delta function 
$\delta_{\mu}(k,f_{\theta}\cdot k)$ 
localizes
the integral to the only point on $\calO_p$ that is invariant under 
rotations, namely the constant supermomentum $p=m-c_2/24$. This 
allows us to replace $k$ by $m-c_2/24$ in the term $e^{i\langle k,\alpha\rangle}$ of eq.~(\ref{pchar}), 
which gives
\be
\langle k,\alpha\rangle
=
\langle p,\alpha\rangle
\refeq{pairing}
\alpha^0(m-c_2/24)
\nn
\ee
where $\alpha^0$ denotes the zeroth Fourier mode of the supertranslation 
$\alpha(\phii)=\sum_{n\in\ZZ}\alpha^ne^{-in\phii}$. None 
of the higher Fourier modes of $\alpha$ contribute to the character. We can thus pull the exponential out of 
the integral in (\ref{pchar}) and the character reduces to
\be
\chi_{m,j}[(f,\alpha)]
=
e^{ij\theta}e^{i\alpha^0(m-c_2/24)}
\int_{\calO_p}d\mu(k)
\delta_{\mu}(k,f_{\theta}\cdot k).
\label{locChar}
\ee
To integrate the delta function, we need to find coordinates on $\calO_p$. As a first step, note that each 
supermomentum
$k(\phii)$ can be expanded in Fourier series as
\be
k(\phii)
=
\sum_{n\in\ZZ}k_ne^{-in\phii},
\label{foufou}
\ee
with $(k_n)^*=k_{-n}$. Then, according to the transformation law (\ref{coadVir}), the action of a 
rotation 
$f_{\theta}(\phii)=\phii+\theta$ on $k$ is simply $\left.\left(f_{\theta}\cdot 
k\right)\right|_{\phii}=k(\phii-\theta)$. 
In terms of Fourier modes, this corresponds to
\be
k_n\mapsto\left[f_{\theta}\cdot k\right]_n=k_ne^{in\theta}.
\nn
\ee
We will soon see that the character obtained using this transformation is divergent, as we 
might expect since the group is infinite-dimensional. To regularize the result, we will therefore consider 
complex rotations instead 
of real ones: let
\be
\tau\equiv\frac{1}{2\pi}(\theta+i\epsilon)
\label{modPar}
\ee
be a complex parameter with $\epsilon>0$, and define the transformation of the supermomentum $k$ under a 
rotation by $2\pi\tau$ to be
\be
k_n\mapsto\left[f_{2\pi\tau}\cdot k\right]_n=
\left\{\begin{array}{lcc}
k_ne^{2\pi in\tau} & \text{if} & n>0,\\
k_0 & \text{if} & n=0,\\
k_ne^{2\pi in\bar\tau} & \text{if} & n<0.\end{array}\right.
\label{tFour}
\ee
We will show below that this seemingly {\it ad hoc} modification is related to thermodynamical considerations 
and to the fact that the $\BMS$ 
group is a ``high-energy'' limit of two Virasoro groups. Note also that this 
prescription allows for ``Euclidean'' rotations (that is, rotations through an imaginary angle) while 
preserving the reality condition $(k_n)^*=k_{-n}$.\\

The question now is how to express the measure $\mu$ and the delta function $\delta_{\mu}$ in terms of modes 
$k_n$. On the orbit $\calO_p$, diffeomorphic to $\Diff/S^1$, each 
supermomentum $k$ is uniquely determined by its non-zero Fourier modes; in other words, the 
non-zero Fourier modes of $k$ determine its zero-mode, given that $k$ belongs to $\calO_p$. An easy way to 
see this locally, in a neighbourhood of $p$, is to consider the action 
(\ref{cake}) of infinitesimal superrotations on $p$ \cite{Witten}. Upon expanding the vector field 
$X(\phii)\frac{d}{d\phii}$ as a Fourier series
\be
X(\phii)
=
i\sum_{n\in\ZZ}X_ne^{-in\phii}
\nn
\ee
and writing $p=m-c_2/24$, this action becomes
\be
\left(\delta_Xp\right)(\phii)
=
\sum_{n\in\ZZ}2n\left(m+\frac{c_2}{24}(n^2-1)\right)X_ne^{-in\phii}
\equiv
\sum_{n\in\ZZ}\delta p_ne^{-in\phii}.
\label{infim}
\ee
In this expression, the zero-mode $\delta p_0$ always vanishes, regardless of $X(\phii)$. By contrast, all 
other Fourier modes $\delta p_n$ can be made non-zero by a suitable choice of $X(\phii)$. Thus, at least in a 
neighbourhood of $p$, we may choose the non-zero Fourier modes of supermomenta as coordinates on 
$\calO_p$. In the notation of (\ref{cake}) and (\ref{foufou}), when $k$ is close to $p$ so 
that $k(\phii)=p+\eps\left(\delta_Xp\right)(\phii)$, the modes $k_n$ with non-zero $n$ reduce to $\eps\delta 
p_n$ (while $k_0$ reduces to $m-c_2/24$ to first order in $\eps$). Thus, in terms of $k_n$'s, the 
supermomentum measure $\mu$ in (\ref{locChar}) must 
take the form
\be
d\mu(k)=\text{(Some $k$-dependent prefactor)}\times\prod_{n\in\ZZ^*}dk_n,
\label{measOp}
\ee
where the prefactor is generally unknown. In the standard notation of quantum mechanics, the infinite product 
$\prod_{n\in\ZZ^*}dk_n$ would be written as a path integral measure $\calD k$, being understood that the 
zero-mode of $k$ must not be integrated over. By virtue of the definition (\ref{testFct}), the delta function 
corresponding to $\mu$ reads
\be
\delta_{\mu}(q,k)
=
\text{(Some $k$-dependent prefactor)}^{-1}\times\prod_{n\in\ZZ^*}\delta(q_n-k_n),
\nn
\ee
where the $\delta$'s on the right-hand side are the usual one-dimensional Dirac delta 
functions. Thus, the prefactors of $d\mu$ and $\delta_{\mu}$ cancel out and 
expression (\ref{locChar}) boils down to
\bea
\chi_{m,j}[(f,\alpha)]
& = &
e^{ij\theta}e^{i\alpha^0(m-c_2/24)}
\int_{\RR^{2\infty}}\prod_{n\in\ZZ^*}dk_n\prod_{n\in\ZZ^*}
\delta\left(k_n-\left[f_{2\pi\tau}\cdot k\right]_n\right)\nn\\
\label{star12}
& \stackrel{\text{(\ref{tFour})}}{=} &
e^{ij\theta}e^{i\alpha^0(m-c_2/24)}
\Big|
\int_{\RR^{\infty}}\prod_{n=1}^{+\infty}dk_n\prod_{n=1}^{+\infty}
\delta\left(k_n(1-e^{2\pi in\tau})\right)
\Big|^2,
\eea
where we replaced the real angle $\theta$ by its complex counterpart $2\pi\tau$, defined by (\ref{modPar}). 
Writing $q\equiv\exp[2\pi i\tau]$ and evaluating the integral, the 
character of a massive $\BMS$ particle finally reduces to
\be
\boxed{\chi_{m,j}[(f,\alpha)]
=
e^{ij\theta}e^{i\alpha^0(m-c_2/24)}
\frac{1}{\prod_{n=1}^{+\infty}|1-q^n|^2}}
\quad\text{if}\;\;\text{Rot}(f)=\theta\neq 0.
\label{charBMS}
\ee
This can be rewritten as
\be
\chi_{m,j}[(f,\alpha)]
=
\frac{|q|^{1/12}}{|\eta(\tau)|^2}
e^{ij\theta}e^{i\alpha^0(m-c_2/24)}
\nn
\ee
in terms of the Dedekind Eta function
\be
\eta(\tau)\equiv q^{1/24}\prod_{n=1}^{+\infty}(1-q^n).
\nn
\ee
We stress that, at this stage, and in contrast to standard conformal field theory, 
the number $\tau$ should not be interpreted as a modular parameter. The small parameter 
$\epsilon$ in 
(\ref{modPar}) was merely introduced to ensure convergence of the determinant arising from the integration 
of the delta function in (\ref{star12}). This being said, the occurrence of the Eta function is compatible 
with the modular transformations used in 
\cite{BarnichEntropy,BagchiEntropy} to compute the entropy of cosmological 
solutions.

\subsection{Comparison to Poincar\'e and Virasoro characters}
\label{subsecCompChar}

\subsubsection*{From Poincar\'e to {\bf $\text{BMS}_3$}}

Expression (\ref{charBMS}) is a natural
extension of the Poincar\'e character (\ref{charrot}). Indeed, taking $\epsilon=0$ in 
(\ref{charBMS}) and being careless about convergence issues, we find
\be
\chi_{m,j}[(f,\alpha)]
=
e^{ij\theta}e^{i\alpha^0(m-c_2/24)}
\prod_{n=1}^{+\infty}\frac{1}{4\sin^2(n\theta/2)}.
\nn
\ee
Up to the normalization of energy, the term $n=1$ exactly reproduces (\ref{charrot}), while the contribution 
of higher Fourier modes can be loosely 
interpreted as coming from the infinitely many Poincar\'e subgroups of $\BMS$.\\

This phenomenon is analogous to the relation between $\SL$ and Virasoro. Indeed, the 
character of a rotation $f_{2\pi\tau}$ through a complex angle $2\pi\tau$ with positive imaginary part in a 
highest weight representation 
of the $\sl(2,\RR)$ algebra is
\be
\chi_h[f_{2\pi\tau}]
=
\text{Tr}\left(q^{L_0}\right)
=
q^h\left(1+q+q^2+\cdots\right)
=
\frac{q^h}{1-q}
\label{SLchar}
\ee
where $q\equiv e^{2\pi i\tau}$ and where $L_0$ denotes the generator of 
rotations in $\sl(2,\RR)$. The term $q^h$ comes from the highest weight state (with weight $h$), while 
$(1-q)^{-1}$ is the contribution of its 
$\sl(2,\RR)$ descendants. This should be compared to the character of a highest weight 
representation of the Virasoro algebra at central charge $c>1$,
\be
\chi_h[f_{2\pi\tau}]
=
\text{Tr}\left(q^{L_0-c/24}\right)
=
\frac{q^{h-c/24}}{\prod_{n=1}^{+\infty}(1-q^n)},
\label{VirChar}
\ee
which is obviously a generalization of (\ref{SLchar}), including a 
contribution 
from higher Fourier modes reminiscent of the infinitely many $\SL$ subgroups of the Virasoro group. Again, 
the denominator is interpreted as the contribution of descendant states.

\subsubsection*{{\bf $\text{BMS}_3$} characters as a flat limit of Virasoro characters}

The 
divergence of the $\BMS$ character (\ref{charBMS}) as $\epsilon\rightarrow0$ is identical to the divergence 
of the Virasoro character (\ref{VirChar}) as $\tau$ becomes real. In this sense, the divergence of the $\BMS$ 
character is not a 
pathology of the $\BMS$ group, but rather a general behaviour we should expect from any infinite-dimensional 
group; the divergence is cured by adding an imaginary part $i\epsilon$ to the rotation angle. The origin 
of this imaginary part can be 
traced back to the fact that the $\BMS$ group is a flat/ultrarelativistic limit of two Virasoro groups 
\cite{BarnichSS,BarnichFlim}, as follows.\\

If we denote by $L_m$ and $\bar L_m$ the 
generators of two 
commuting copies of the Virasoro algebra with central charges $c$ and $\bar c$ and if $\ell$ is some length 
scale, one can 
define
\be
P_n\equiv\frac{1}{\ell}\left(L_n+\bar L_{-n}\right),
\quad
J_n\equiv L_n-\bar L_{-n},
\quad
c_1\equiv c-\bar c
\quad\text{and}\quad
c_2\equiv\frac{c+\bar c}{\ell}.
\label{VirtoBMS}
\ee
In the limit $\ell\rightarrow+\infty$, and provided 
$c_1$ and $c_2$ are 
finite in that limit, these generators span a 
centrally extended 
$\bms$ algebra:
\be
[J_n,J_m]=(n-m)J_{n+m}+\frac{c_1}{12}n^3\delta_{n+m,0},\;\;
[J_n,P_m]=(n-m)P_{n+m}+\frac{c_2}{12}n^3\delta_{n+m,0},\;\;
[P_n,P_m]=0.
\label{bms}
\ee
In the context of three-dimensional Einstein gravity on AdS with Brown-Henneaux boundary conditions 
\cite{BrownHenneaux}, the $L_n$'s and the $\bar L_n$'s are surface charges associated with asymptotic 
symmetries. The parameter $\ell$ is related to the negative cosmological constant 
$\Lambda=-1/\ell^2$, and $c=\bar c=3\ell/2G$. The regime of large $\ell$ then corresponds to the ``flat 
limit'' of AdS, and the $\bms$ central charges are finite since $c_2=3/G$ and $c_1=0$.\\

Now, the 
character of a highest weight representation of two copies of the Virasoro algebra, labelled by the 
(strictly positive) highest weights $h$, $\bar h$ and the central charges $c>1$ and $\bar c>1$, is
\be
\text{Tr}\left(q^{L_0-c/24}\bar q^{\bar L_0-\bar c/24}\right)
\refeq{VirChar}
\frac{q^{h-c/24}\bar q^{\bar h-\bar c/24}}{\prod_{n=1}^{+\infty}|1-q^n|^2},
\quad q=e^{2\pi i\tau}.
\label{VirCharBis}
\ee
Let us see how we can recover the $\BMS$ character (\ref{charBMS}) as a flat limit of this expression. First, 
we write the modular parameter as $\tau=\frac{1}{2\pi}(\theta+i\beta/\ell)$, where $\beta>0$ is 
$\ell$-independent. Provided the highest weights $h$ and $\bar h$ scale 
with $\ell$ in such a way that the numbers
\be
m\equiv\lim_{\ell\rightarrow+\infty}\frac{1}{\ell}(h+\bar h)
\quad\text{and}\quad
j\equiv\lim_{\ell\rightarrow+\infty}(h-\bar h)-\frac{c_1}{24}
\label{mj}
\ee
be finite, the large $\ell$ limit of the quantities 
appearing on the right-hand side of (\ref{VirCharBis}) is
\be
\tau\sim\frac{1}{2\pi}(\theta+i\epsilon),
\quad
q^{h-c/24}\bar q^{\bar h-\bar c/24}\sim e^{ij\theta}e^{-\beta(m-c_2/24)}.
\nn
\ee
(Here the imaginary part of $\tau$ goes to zero, but we keep writing it as $\epsilon>0$ to reproduce the 
regularization used in (\ref{charBMS}).) We conclude that the flat limit of 
(\ref{VirCharBis}) is
\be
\lim_{\ell\rightarrow+\infty}\text{Tr}\left(q^{L_0-c/24}\bar q^{\bar L_0-c/24}\right)
=
e^{ij\theta}e^{-\beta(m-c_2/24)}
\frac{1}{\prod_{n=1}^{+\infty}|1-q^n|^2},
\nn
\ee
which coincides with the $\BMS$ character (\ref{charBMS}) provided we consider a supertranslation whose 
zero-mode is a Euclidean time translation, $\alpha=i\beta$. The 
left-hand side of this expression can be 
interpreted as
\be
\lim_{\ell\rightarrow+\infty}\text{Tr}\left(q^{L_0-c/24}\bar q^{\bar L_0-c/24}\right)
\stackrel{\text{(\ref{VirtoBMS})}}{=}
\text{Tr}\left(e^{i\theta J_0}e^{-\beta(P_0-c_2/24)}\right)
=
\chi[(f_{\theta},\alpha)],
\nn
\ee
where $f_{\theta}$ denotes a rotation by $\theta$. In this form, the matching between the flat limit of the 
Virasoro character (\ref{VirChar}) 
and the $\BMS$ character (\ref{charBMS}) is obvious.\\

By the way, the definition (\ref{mj}) explains why induced representations of $\BMS$ can roughly be seen 
as 
an ultrarelativistic/high-energy limit of Virasoro highest weight representations: keeping $m$ finite and 
non-zero as $\ell$ goes to infinity requires $h$ and/or $\bar h$ to grow linearly with $\ell$, meaning that 
$h+\bar h$ must take an infinitely large value in the limit $\ell\rightarrow+\infty$. This observation, 
together with (\ref{VirtoBMS}), also gives us an intuitive picture of why the energy spectrum of $\BMS$ 
particles is continuous: the typical distance between two consecutive eigenvalues of $P_0=(L_0+\bar 
L_0)/\ell$ is $1/\ell$, which shrinks to zero when $\ell$ goes to infinity. Thus, the only way for $P_0$ to 
have more than one eigenvalue is that its spectrum be continuous. This is by no means a pathology, as the 
same situation occurs with the energy spectrum $q_0=\sqrt{m^2+{\bf q}^2}$ in induced representations of the 
Poincar\'e group.

\subsubsection*{Further comments}

Even though $\BMS$ and Virasoro characters are related through the flat/ultrarelativistic limit just 
described, they are strikingly different in many respects. Indeed, the result (\ref{charBMS}) holds for any 
(positive) value of the central charge $c_2$, any (positive) value of the mass $m$, and any spin $j$. By 
contrast, the character of an irreducible, unitary highest weight representation of the Virasoro 
algebra, labelled by the values of $c$ and $h$, depends heavily on those values: when $c<1$, only certain 
discrete values of $c$ and $h$ lead to unitary representations \cite{DiFran,Blum}, and the corresponding 
character is strikingly different from (\ref{VirChar}) 
\cite{FeiginFuchs01,Wassermann01,Wassermann02}. From that viewpoint, induced representations of 
the 
$\BMS$ group are less intricate than highest weight representations of the Virasoro algebra. Since the former 
are 
high-energy, high central charge limits of 
the latter, this seems reasonable: all complications occurring at small $c$ vanish when $\ell$ goes to 
infinity, since $c$ scales linearly with $\ell$ by assumption.\\

We could have guessed that such a simplification would occur on the basis of dimensional arguments. Indeed, 
the parameters labelling induced 
representations of $\BMS$ are dimensionful: both $m$ and $c_2$ have 
dimension of mass. This is manifest in the flat limit relations (\ref{VirtoBMS}) and (\ref{mj}), 
where the mass dimension arises due to the length scale $\ell$. Alternatively, it may be seen as a 
consequence of the fact that supermomenta are an infinite-dimensional generalization of momenta, hence 
naturally carrying a dimension of mass. Thus the values of $m$ and $c_2$ can be tuned at will by a suitable 
choice of units, and, in 
contrast to Virasoro highest weight representations, we should not expect to find sharp 
bifurcations in the structure of $\BMS$ induced representations as $m$ and $c_2$ vary. In this 
sense, the character (\ref{charBMS}) is universal in the region of positive $m$ and $c_2$.

\subsection{Character of the {\bf $\text{BMS}_3$} vacuum representation}
\label{ssecVac}

We now turn to the character of the induced representation with vanishing spin based on the orbit of the 
vacuum, $p=-c_2/24$. 
The computation is very similar to that of subsection \ref{charrBMS}, save for the fact that the little 
group is the Lorentz group $\PSL=\SL/\ZZ_2$ rather than $\mathrm{U}(1)$. In particular, the orbit 
$\calO_{\text{vac}}$ of $p$ is diffeomorphic to the universal Teichm\"uller space $\Diff/\PSL$.\\

As before, the quantity we wish to compute is $\chi[(f,\alpha)]$, where $\alpha$ is any supertranslation. 
Since the little group here is larger than $\mathrm{U}(1)$, we can obtain a non-trivial character even when 
$f$ is not conjugate to a rotation $-$ for example if $f$ is a boost in $\PSL$. Although this computation may 
be interesting for physics, we will not consider it here and we will focus on the case in which $f$ is 
conjugate to a rotation $f_{\theta}$. We will also assume, as in subsection \ref{charrBMS}, that the rotation 
number $\theta$ of $f$, given by (\ref{BigStar}), is non-zero. The delta 
function in the Frobenius formula (\ref{Frobis}) then  forces the integral to pick its only non-zero 
contribution 
from the unique rotation-invariant point on the orbit $\calO_{\text{vac}}$, that is, the point $p=-c_2/24$. 
The character can thus be written in a form analogous to (\ref{locChar}),
\be
\chi_{\text{vac}}[(f,\alpha)]
=
e^{-i\alpha^0c_2/24}
\int_{\calO_{\text{vac}}}d\mu(k)
\delta_{\mu}(k,f_{\theta}\cdot k),
\label{vacChar}
\ee
where $\mu$ is now a quasi-invariant measure on $\calO_{\text{vac}}\cong\Diff/\PSL$. We can of course expand 
any supermomentum $k$ belonging to $\calO_{\text{vac}}$ 
as a Fourier series (\ref{foufou}), and we define the transformation of Fourier modes under complex rotations 
to be (\ref{tFour}). The subtlety now is that we must understand which Fourier modes appear in the measure, 
or equivalently which modes can be used as coordinates on the orbit.\\

Fortunately, we only need local coordinates in a neighbourhood of the rotation-invariant point, 
$p=-c_2/24$. We can thus rely again on an argument based on the action (\ref{infim}) of infinitesimal 
superrotations on constant supermomenta. Taking $m=0$ in that expression, we now find that not only the 
zero-mode $\delta p_0$, but also the modes $\delta p_{\pm1}$, vanish for all choices of $X(\phii)$. This 
means that, in a neighbourhood of $p=-c_2/24$, we can endow the orbit $\calO_{\text{vac}}$ with coordinates 
given by the 
higher Fourier modes of $k(\phii)$, that is, the modes $k_{\pm2}$, $k_{\pm3}$, {\it etc}. In particular, in 
this neighbourhood of the rotation-invariant point, the 
measure $\mu$ on $\calO_{\text{vac}}$ takes the form
\be
d\mu(k)=\text{(Some $k$-dependent prefactor)}\times\prod_{n=2}^{+\infty}dk_ndk_{-n},
\nn
\ee
where, as in (\ref{measOp}), the prefactor is unknown. Once more, the definition 
(\ref{testFct}) of the corresponding delta function ensures that this prefactor cancels with its 
inverse appearing in $\delta_{\mu}$, which reduces the vacuum character (\ref{vacChar}) to
\bea
\chi_{\text{vac}}[(f,\alpha)]
& = &
e^{-i\alpha^0c_2/24}
\int_{\RR^{2\infty-2}}\prod_{n=2}^{+\infty}dk_ndk_{-n}\prod_{n=2}^{+\infty}
\delta(k_n-[f_{2\pi\tau}\cdot k]_n)
\delta(k_{-n}-[f_{2\pi\tau}\cdot k]_{-n})\nn\\
& \stackrel{\text{(\ref{tFour})}}{=} &
e^{-i\alpha^0c_2/24}
\Big|
\int_{\RR^{\infty-1}}\prod_{n=2}^{+\infty}dk_n\prod_{n=2}^{+\infty}
\delta\left(k_n(1-q^n)\right)
\Big|^2,
\nn
\eea
where $q\equiv\exp[2\pi i\tau]$ and $\tau=(\theta+i\epsilon)/2\pi$. Integrating the delta functions and 
taking into account the determinant, we end up with
\be
\boxed{\chi_{\text{vac}}[(f,\alpha)]
=
e^{-i\alpha^0c_2/24}
\frac{1}{\prod_{n=2}^{+\infty}|1-q^n|^2}}
\quad\text{when}\;\;\text{Rot}(f)=\theta\neq0.
\label{charBMSvac}
\ee
As before, this expression can be interpreted as the trace
\be
\text{Tr}\left(e^{i\theta J_0}e^{i\alpha^0(P_0-c_2/24)}\right)
\nn
\ee
in the vacuum representation of $\BMS$. It can also be recovered as a flat limit of the product of 
two vacuum Virasoro characters. In that context, the truncated product 
$\prod_{n=2}^{+\infty}(...)$ arises due to $\SL$-invariance of the Virasoro vacuum.

\section*{Conclusion and outlook}
\addcontentsline{toc}{section}{Conclusion and outlook}

In this work we have shown that the standard Frobenius formula (\ref{Frobis}) for characters of induced 
representations can be applied to the $\BMS$ group. Our key results were formulas 
(\ref{charBMS}) and (\ref{charBMSvac}), which can also be seen as flat limits of the corresponding Virasoro 
characters. A crucial step during the computation was the use of local coordinates on Virasoro orbits, which 
allowed us to write down the integration measure up to an unknown prefactor. The latter eventually turned out 
to be irrelevant, as it was cancelled by the prefactor of the delta function.\\

One important case is still missing in our considerations. Indeed, in subsection \ref{charrBMS}, we 
restricted our attention to superrotations conjugate to non-trivial rigid rotations of the circle. In doing 
so, we left aside the computation of $\BMS$ characters associated with pure supertranslations. As already 
pointed out, such characters are necessarily infrared-divergent, as is the Poincar\'e 
character (\ref{timeChar}). If we assume that the $\BMS$ version of this 
divergence can be put under control, and if we focus on pure time translations, we are left with the task of 
computing the $\BMS$ analogue of the integral in (\ref{timeInt}), which is 
schematically of the form
\be
\int_{\calO_p}\calD k\exp\left[itE[k]\right]
\quad\text{or}\quad
\int_{\calO_p}\calD k\exp\left[-\beta E[k]\right].
\label{intint}
\ee
Here $E[k]$ denotes the energy functional on $\calO_p$, 
that is, the zero-mode 
of $k(\phii)$ when $k$ belongs to $\calO_p$. (The second expression in (\ref{intint}) is the Euclidean 
version of the first one.) We did not attempt 
to evaluate this integral here, but hope to address this question in the future.\\

Many other open issues were left aside in the present paper. One of these is the relation of $\BMS$ 
characters 
to one-loop partition functions of three-dimensional asymptotically flat 
gravity, which is analogous to the matching between Virasoro characters and partition functions on 
AdS$_3$ \cite{Giombi}. Indeed, it is easily verified that the flat limit of the heat kernel computation of 
\cite{Giombi} precisely reproduces (\ref{charBMSvac}) as the one-loop partition function of gravitons around 
thermal flat space (provided we identify $\alpha^0=i\beta$). We will turn to a thorough investigation of this 
relation elsewhere. Independently of this, several 
natural extensions of our considerations are available: one may imagine computing characters of 
certain higher-spin extensions of $\BMS$ \cite{Afshar,Grumiller,Matulich} or of the supersymmetric $\BMS$ 
group \cite{Mandal,BarnichDonnay}. In all those cases, the semi-direct product structure of $\BMS$ should be 
essential in determining the appropriate unitary representations and the associated character formula.

\section*{Acknowledgements}

My warmest thanks go to G.~Barnich and H.~Gonz\'alez for advice, support and collaboration on 
related projects, as well as many enlightening discussions on physics in general. I am also grateful to 
D.~Pickrell for pointing me to the references \cite{Shavgulidze,Bogachev} on quasi-invariant 
functional measures, and to K.H.~Neeb for helpful correspondence on the same topic. This work was 
supported by the Fonds de la Recherche Scientifique-FNRS under grant number FC-95570.


\section*{References}
\addcontentsline{toc}{section}{References}

\renewcommand{\section}[2]{}%
 
\def\cprime{$'$}

\providecommand{\href}[2]{#2}\begingroup\raggedright\endgroup

\end{document}